\documentclass[letterpaper,12pt]{article}
\usepackage{graphicx} 
\usepackage{comment}
\usepackage{amsmath}
\usepackage{subcaption}
\usepackage{longtable}
\usepackage{makecell}
\usepackage{authblk}  
\usepackage{siunitx}
\DeclareSIUnit\angstrom{\text{\AA}}

\usepackage[utf8]{inputenc}
\usepackage[english]{babel}
\usepackage{placeins}
\usepackage{csquotes}

\usepackage{geometry} 
\usepackage{chemformula}
\usepackage{booktabs}

\usepackage[style=ieee, url=false]{biblatex}
\usepackage{microtype} 
\usepackage{color}
\definecolor{deepblue}{rgb}{0,0,0.5}
\definecolor{deepred}{rgb}{0.6,0,0}
\definecolor{deepgreen}{rgb}{0,0.5,0}
\definecolor{citeGreen}{RGB}{102,204,10}

\usepackage{hyperref}
\definecolor{mypurple}{RGB}{140,54,140}
\hypersetup{
    colorlinks=true,
    linkcolor=blue,
    filecolor=mypurple,      
    urlcolor=teal,
    citecolor = citeGreen
}

\title{Stripe Antiferromagnetic Ground-State Configuration of FeSe Revealed by Density Functional Theory}
\author[1,*]{Luke Allen Myers}
\author[1]{Nigel Lee En Hew}
\author[1]{Shun-Li Shang}
\author[1]{Zi-Kui Liu}
\affil[1]{Department of Materials Science and Engineering, The Pennsylvania State University, University Park, Pennsylvania 16802, USA}
\affil[*]{Corresponding authors: lam7027@psu.edu}
\date{}

\begin{document}

\maketitle

\section*{Abstract}

The magnetic ground-state configuration of iron selenide (\ch{FeSe}) has been a topic of debate, with experimental evidence suggesting the stripe spin fluctuations as predominant at low temperatures, while density functional theory (DFT) calculations using exchange-correlation (XC) functional of the Generalized Gradient Approximation (GGA) have historically predicted the antiferromagnetic (AFM) dimer configuration. In this study, we utilize the r\textsuperscript{2}SCAN functional, a variant of the Strongly Constrained and Appropriately Normed (SCAN) meta-GGA, to investigate the magnetic configurations of \ch{FeSe}. It is found that r\textsuperscript{2}SCAN predicts a stripe-AFM ground-state configuration with an anti-parallel spin alignment between layers. The energy difference between the parallel and anti-parallel inter-planar spin alignments is approximately 1.7 meV/atom, predicting a significant but previously unreported interlayer spin coupling not yet observed by experiments. The present study underscores the importance of accurate XC functionals, such as r\textsuperscript{2}SCAN, in predicting the magnetic ground-state configuration of complex materials like \ch{FeSe}, highlighting its potential to predict magnetic interactions more reliably than traditional GGA functionals by adhering to exact constraints.

\section{Introduction}

FeSe and its magnetic ground state have been intensely studied due to its unusual magnetism and highly tunable superconductivity \cite{Bohmer2018NematicityFeSe}. Although the superconducting transition temperature ($T_c$) at ambient pressure is 8 K, the application of hydrostatic pressure can raise $T_c$ to 34 K \cite{Hsu2008Superconductivity-FeSe, Miyoshi2014EnhancedPressure}. Superconductivity in FeSe can be further enhanced above 100 K for a single layer epitaxially grown on a \ch{SrTiO3} substrate \cite{Ge2015SuperconductivitySrTiO3} possibly due to interface-enhanced electron-phonon coupling \cite{Ge2015SuperconductivitySrTiO3, Wang2012Interface-InducedSrTiO3, Ginzburg1964OnSuperconductivity, Cohen1967SuperconductiveBarriers, Strongin1968EnhancedFilms, Lee2014InterfacialSrTiO3, Coh2015LargeMonolayer}. With only a few exceptions, antiferromagnetism in iron-based compounds is typically stripe-antiferromagnetic (AFM) ordered \cite{Bohmer2018NematicityFeSe} with in-plane alignment antiparallel in one direction and parallel in the perpendicular direction. Since the stripe-AFM ordering breaks the \textit{C4} rotational symmetry, the paramagnetic (PM) to AFM phase transition typically coincides with a tetragonal to orthorhombic phase transition, and it has been argued that the structural transition is a consequence of stripe magnetic order \cite{Dai2015AntiferromagneticSuperconductors} However, FeSe undergoes a tetragonal to orthorhombic phase transition at about 90 K \cite{Kothapalli2016StrongFeSe, Margadonna2008CrystalSuperconductorw, McQueen2009Tetragonal-to-orthorhombicFe1.01Se}, but magnetic ordering remains undetected by low-temperature Mössbauer spectroscopy \cite{McQueen2009Tetragonal-to-orthorhombicFe1.01Se}, magnetotransport measurements \cite{Sun2016Dome-shapedFeSe}, nuclear magnetic resonance \cite{Baek2014Orbital-drivenFeSe, Bohmer2015OriginNematicity,  Imai2009WhyPressure}, and neutron scattering \cite{Wang2015StrongFeSe, Wang2016MagneticFeSe}. While the explanatory mechanism for the persistent paramagnetism remains unknown, theoretical studies have hypothesized it to be a consequence of magnetic frustration \cite{Wang2015NematicityFeSe, Glasbrenner2015EffectChalcogenides, Yu2015AntiferroquadrupolarFeSe, Cao2015AntiferromagneticFeSe, Chubukov2015OriginFeSe}.

\begin{figure*}[h]
    \centering
    \begin{subfigure}[t]{0.125\textwidth}
        \centering
        \includegraphics[width=1\linewidth]{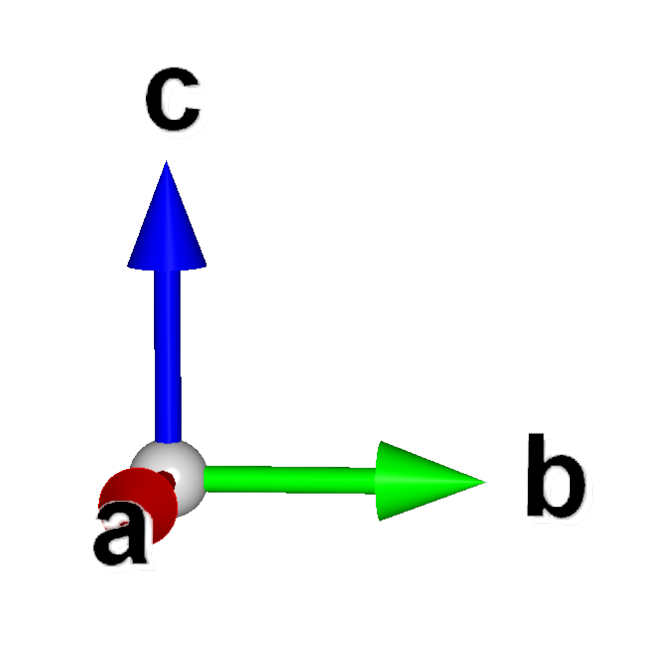}
    \end{subfigure}%
    ~ 
    \centering
    \begin{subfigure}[t]{0.425\textwidth}
        \centering
        \includegraphics[width=1\linewidth]{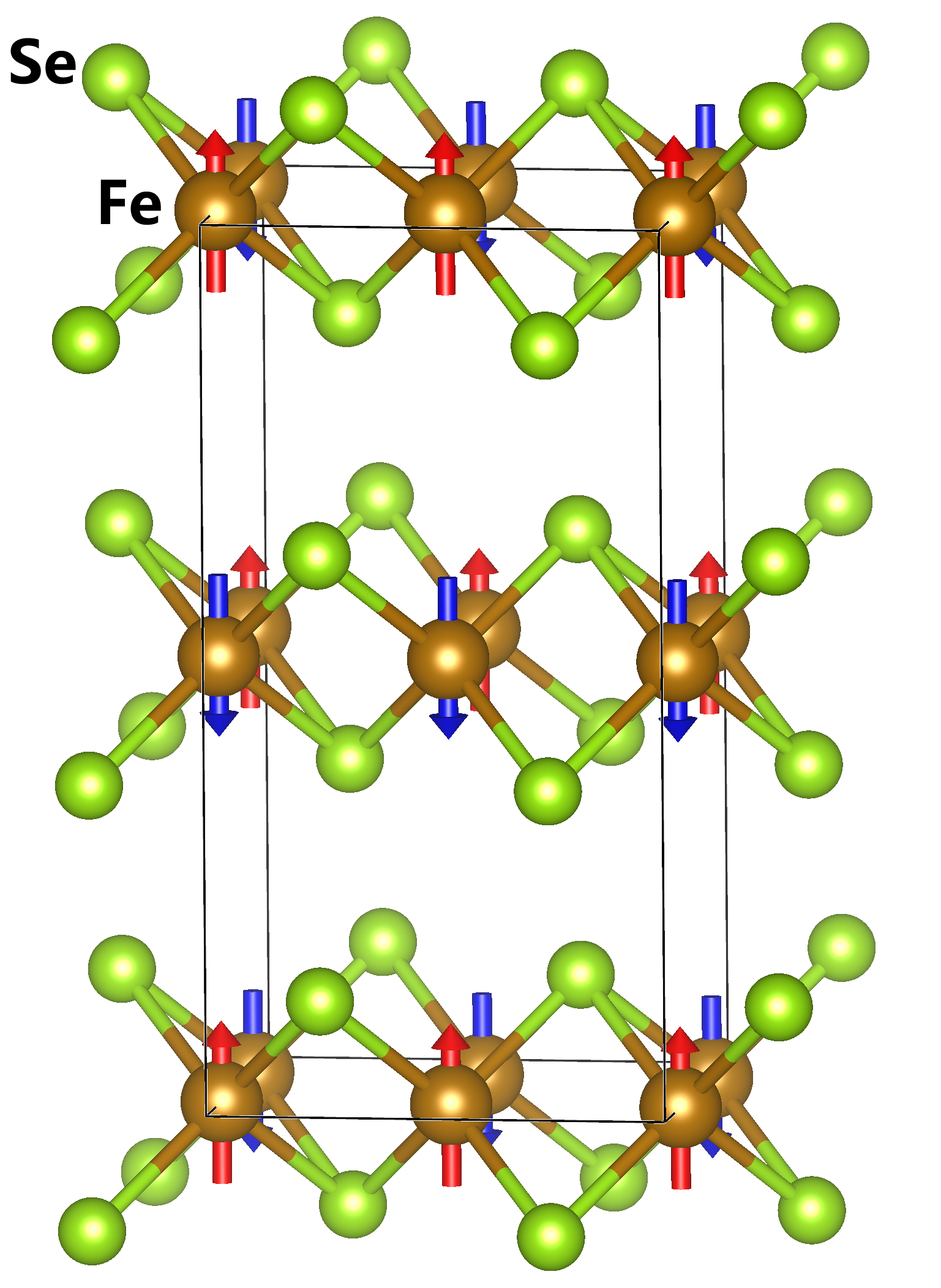}
        \caption{r\textsuperscript{2}SCAN ground-state configuration}
        \label{fig:r2scan_ground}
    \end{subfigure}%
    ~ 
    \begin{subfigure}[t]{0.425\textwidth}
        \centering
        \includegraphics[width=1\linewidth]{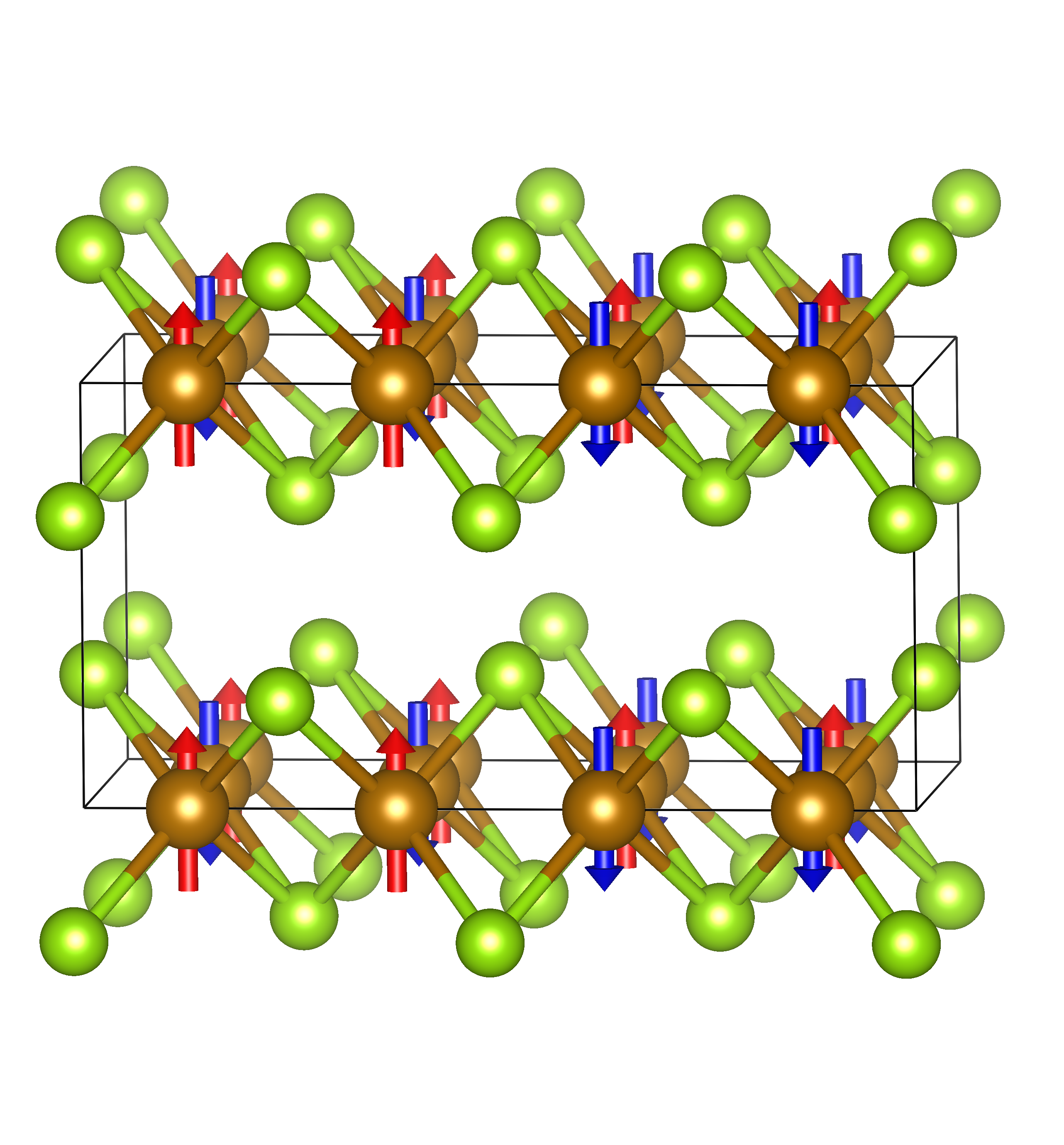}
        \caption{GGA-PBE ground-state configuration }
        \label{fig:pbe_ground}
    \end{subfigure}
    \caption{16-atom supercells of FeSe for (a) the Stripe-0 r\textsuperscript{2}SCAN ground-state configuration and (b) the dimer GGA-PBE ground-state configuration. Fe (Se) atoms are shown in brown (green). Arrows denote relative spin directions, as spin-orbit coupling was not included.}
\end{figure*}

In line with the parent compounds of mani iron-based superconductors \cite{Fernandes2014WhatIron-basedsuperconductors, DeLaCruz2008MagneticSystems, Rotter2008SuperconductivityBa1-xKxFe2As2, Torikachvili2008PressureCaFe2As2, Dai2012MagnetismSuperconductors}, early theoretical studies on FeSe considered stripe magnetic ordering \cite{Louca2010LocalTex, Subedi2008DensitySuperconductivity, Bazhirov2013EffectsMonolayer, Ma2009First-principlesOrder, Cao2014InterfacialFilms} (Figure \ref{fig:r2scan_ground}) to be the most stable. However, Cao et al. \cite{Cao2015AntiferromagneticFeSe} and the subsequent density functional theory (DFT) studies \cite{Glasbrenner2015EffectChalcogenides, Liu2016NematicFeSe} found that the dimer (or sometimes pair-checkerboard) order (Figure \ref{fig:pbe_ground}) has a lower energy than the stripe order for FeSe \cite{Cao2015AntiferromagneticFeSe, Glasbrenner2015EffectChalcogenides, Liu2016NematicFeSe}. This is, however, in contrast with neutron-scattering experiments, which show substantial stripe spin fluctuations of wavevector $Q=(\pi,0)$ \cite{Wang2016MagneticFeSe, Wang2015StrongFeSe, Rahn2015StrongSpectroscopy}. Atypical magnetic ground states have also been proposed, including nematic quantum paramagnetism \cite{Wang2015NematicityFeSe}, antiferroquadrupolar order \cite{Yu2015AntiferroquadrupolarFeSe}, and ferroquadrupolar order \cite{Wang2016SpinFeSe}.

The generalized gradient approximation (GGA) developed by Perdew, Burke, and Ernzerhof (GGA-PBE) \cite{Perdew1996GeneralizedSimple} is one of the most used and well-known exchange-correlation (XC) functionals. However, the recent development of the Strongly Constrained and Appropriately Normed (SCAN) meta-GGA \cite{Sun2015StronglyFunctional} has shown improved accuracy for predicting ground states, formation enthalpies, and other material properties compared to GGA-PBE  \cite{Zhang2018EfficientAccuracy, Bartel2019TheStability, Sun2017ThermodynamicNitrides, Park2021ComparisonProperties, Yang2019RationalizingFunctional}. Restored-regularized SCAN (r2SCAN) \cite{Furness2020AccurateApproximation} is a variant that retains the accuracy of SCAN while improving the calculation efficiency. The basis for the predictive success of SCAN and r\textsuperscript{2}SCAN is the satisfaction of all 17 known exact constraints for a meta-GGA, making it genuinely predictive rather than fitting to empirical parameters \cite{Kaplan2023TheTheory}. Advancements have also been made with regard to van der Waals (vdW) dispersion corrections, which are necessary for a proper description of FeSe due to its layered structure. The vdW dispersion correction functionals can be classified as either local or nonlocal, most notably the nonlocal revised Vydrov–van Voorhis (rVV10) \cite{Sabatini2013NonlocalEfficient}, which was parameterized for r2SCAN by Ning et al. \cite{Ning2022WorkhorseR2SCAN+rVV10} and for GGA-PBE by Peng and Perdew \cite{Peng2017RehabilitationMaterials} for layered materials (and named rVV10L).

In the present work, we show an extensive comparison of possible spin configurations and compare the results for vdW dispersion-corrected r\textsuperscript{2}SCAN and GGA-PBE functionals. For precise volume and energy in the relaxed state, we calculate energy-volume (E-V) curves for the lower energy configurations. The present work suggests a stripe-AFM magnetic ordering for the ground-state configuration of FeSe as calculated by r\textsuperscript{2}SCAN with the rVV10 vdW dispersion correction.

\section{Computational Methods}

\subsection{Generating Symmetry-Broken Structures}
Assuming collinear spin magnetism, the spin associated with each \ch{Fe} atom can either be up or down. For a given supercell of FeSe with $n$ \ch{Fe} atoms, there are, therefore, $2^n$ possible magnetic spin arrangements. However, many of these arrangements are symmetrically equivalent. For example, when ignoring spin-orbit coupling, an arrangement with all spin-up is symmetrically equivalent to all spin-down. In addition to the magnetic arrangements within a single supercell, multiple unique supercells can be made for a given number of formula units. Symmetry analysis was performed using the \texttt{genstr} command of the Alloy Theoretic Automated Toolkit (ATAT) \cite{vandeWalle2002TheGuide} to generate all unique symmetry-broken magnetic structures of \ch{FeSe} for $n$-formula units (f.u.) or fewer. Considering the computational resource burden and the rapidly increase number of unique symmetry-broken structures for increasing $n$, the present work limited the scope of symmetry-broken structures to 8-formula units (16 atoms) or less of FeSe, which generates 33 unique supercells and 383 unique magnetic structures for the Cmme crystallographic space group. Structure files for the eight lowest energy structures and other named structures are given in the Ancillary files.

\subsection{Details of First-Principles Calculations}
In the present work, all DFT-based first-principles calculations were performed using the Vienna Ab initio Simulation Package (VASP) \cite{Kresse1996EfficientSet, Kresse1996EfficiencySet}. The projector augmented wave (PAW) method was used to calculate the ion-electron interaction \cite{Blochl1994ProjectorMethod, Kresse1999FromMethod}.
A single fixed volume calculation was performed for each symmetry-broken structure using the r\textsuperscript{2}SCAN + rVV10 density functional approximation (DFA) to filter out high-energy configurations at reduced computational cost. Both cell shape and atomic positions were relaxed while cell volume was kept fixed (i.e., using the VASP setting of \texttt{ISIF = 4}). The volume chosen was 21.5 $\text{\AA}^3/\text{atom}$, which is near the volume of the fully relaxed Stripe-AFM structure. The energy cutoff (ENCUT) for the plane-wave basis set was 520 eV. The global convergence criterion for the electronic self-consistency loop (SC-loop, i.e., the EDIFF) was set to a system energy difference of $2 \times 10^{-4} \ \si{eV}$, which is sufficient to identify the lower energy configurations. The k-point mesh for each symmetry-broken structure was generated using pymatgen \cite{Ong2013PythonAnalysis} with a k-point density of 6000 k-points per reciprocal atom (KPPA), which corresponds to a $9\times9\times9$ mesh for the 8-atom FeSe unit cell with space group Cmme. The Gaussian smearing technique with a width of 0.05 eV was used for the partial occupancies. Additionally, non-spherical contributions related to the gradient of the density in the PAW spheres were included (i.e., the VASP setting, \texttt{LASPH = True}). The conjugate gradient algorithm was used to update atomic positions between SC-loops. A high value of $\pm 5 \ \mu_B$ was used to initialize the Fe magnetic moments. The key VASP settings are summarized in Table \ref{tab:vasp_settings} in the Appendix.

The eight lowest energy configurations from the single fixed volume calculations were then selected for calculation of the E-V curves to find the precise minimum energy. Each point on the E-V curves was calculated successively, starting with the largest volume, by scaling the relaxed atomic positions to a slightly smaller volume and then restarting the calculations using the charge density of the previous (larger) volume as a starting point. Additionally each point was calculated using a three-step procedure:
\begin{enumerate}
    \item Relax with fixed volume and free lattice parameters and atomic positions,
    \item Repeat the relaxation restarting with the results of step 1,
    \item Calculate the final energy for that volume using the tetrahedron method with Blöchl corrections \cite{Blochl1994ImprovedIntegrations}.
\end{enumerate}
This process was performed for 15 linearly spaced volumes starting with $\SI{23.125}{\angstrom^3\per atom}$ and ending with $\SI{18.125}{\angstrom^3\per atom}$. This procedure enables precise energy calculations and helps to maintain the same magnetic configuration from high volume to low volume during the E-V curve calculations. For precise relaxation energies, the global convergence criterion for the electronic SC-loop and the k-point density were increased to a system energy of $5 \times 10^{-7} \ \si{eV}$ and 8000 KPPA, respectively. The minimum energy of each curve was calculated by fitting the four-parameter Birch--Murnaghan equation of state (EOS) \cite{Birch1947FiniteCrystals, Birch1978Finite300K}. Additionally, a single E-V curve was calculated for the Dimer configuration using the PBE-GGA + rvv10L DFA, which is presented in Figure \ref{fig:pbe_rvv10l_dimer_ev} in the Appendix.

\section{Results and Discussion}

The ground-state configuration of FeSe as predicted by r\textsuperscript{2}SCAN is stripe-AFM with an anti-parallel out-of-plane ordering as shown in Figure \ref{fig:r2scan_ground}. This result is in contrast to the results from GGA-PBE-based results in the present work and in the literature, which find that staggered dimer is the magnetic configuration (Figure \ref{fig:pbe_ground}) of the ground state \cite{Cao2015AntiferromagneticFeSe, Glasbrenner2015EffectChalcogenides, Liu2016NematicFeSe}. Inelastic neutron-scattering experiments, however, showed that the stripe order with wavevector $Q = (\pi,0)$ is dominant at 4 K \cite{Wang2016MagneticFeSe}. These results highlight the importance of accurate DFAs when predicting the ground state of materials.

\begin{figure}
    \centering
    \includegraphics[width=0.5\linewidth]{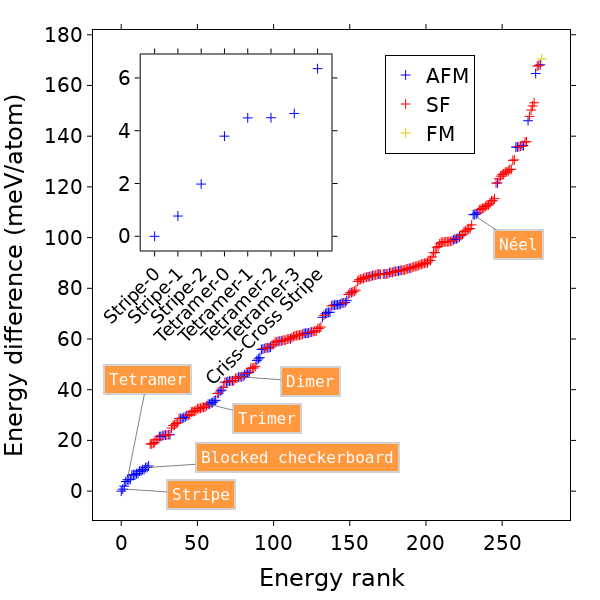}
    \caption{Calculated energy difference for each configuration relative to the lowest energy configuration (stripe-0) by the r\textsuperscript{2}SCAN + rVV10 DFA at a volume of $\SI{21.5}{\angstrom^3\per atom}$. The inset shows the energy difference for the eight lowest configurations. A data table for this figure can be found in the Ancillary files.}
    \label{fig:r2scan_config_energy}
\end{figure}

\begin{figure*}
    \centering
    \begin{subfigure}[t]{0.14\textwidth}
        \centering
        \includegraphics[width=1\linewidth]{compass.png}
    \end{subfigure}%
    ~
    \begin{subfigure}[t]{0.275\textwidth}
        \centering
        \includegraphics[width=1\linewidth]{stripe-0.png}
        \caption{Stripe-0}
        \label{fig:stripe-0}
    \end{subfigure}%
    ~ 
    \begin{subfigure}[t]{0.275\textwidth}
        \centering
        \includegraphics[width=1\linewidth]{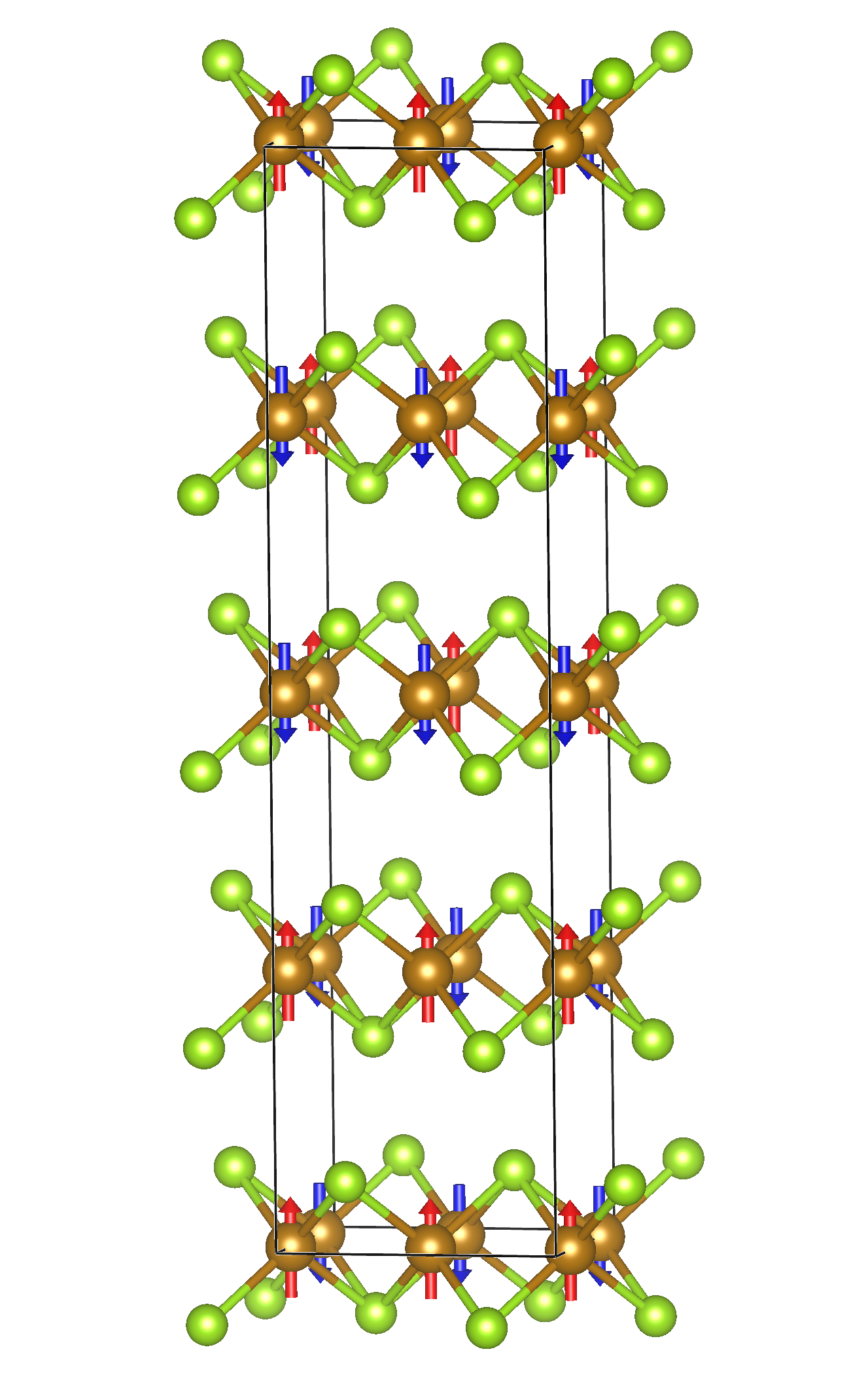}
        \caption{Stripe-1}
        \label{fig:stripe-1}
    \end{subfigure}%
    ~ 
    \begin{subfigure}[t]{0.275\textwidth}
        \centering
        \includegraphics[width=1\linewidth]{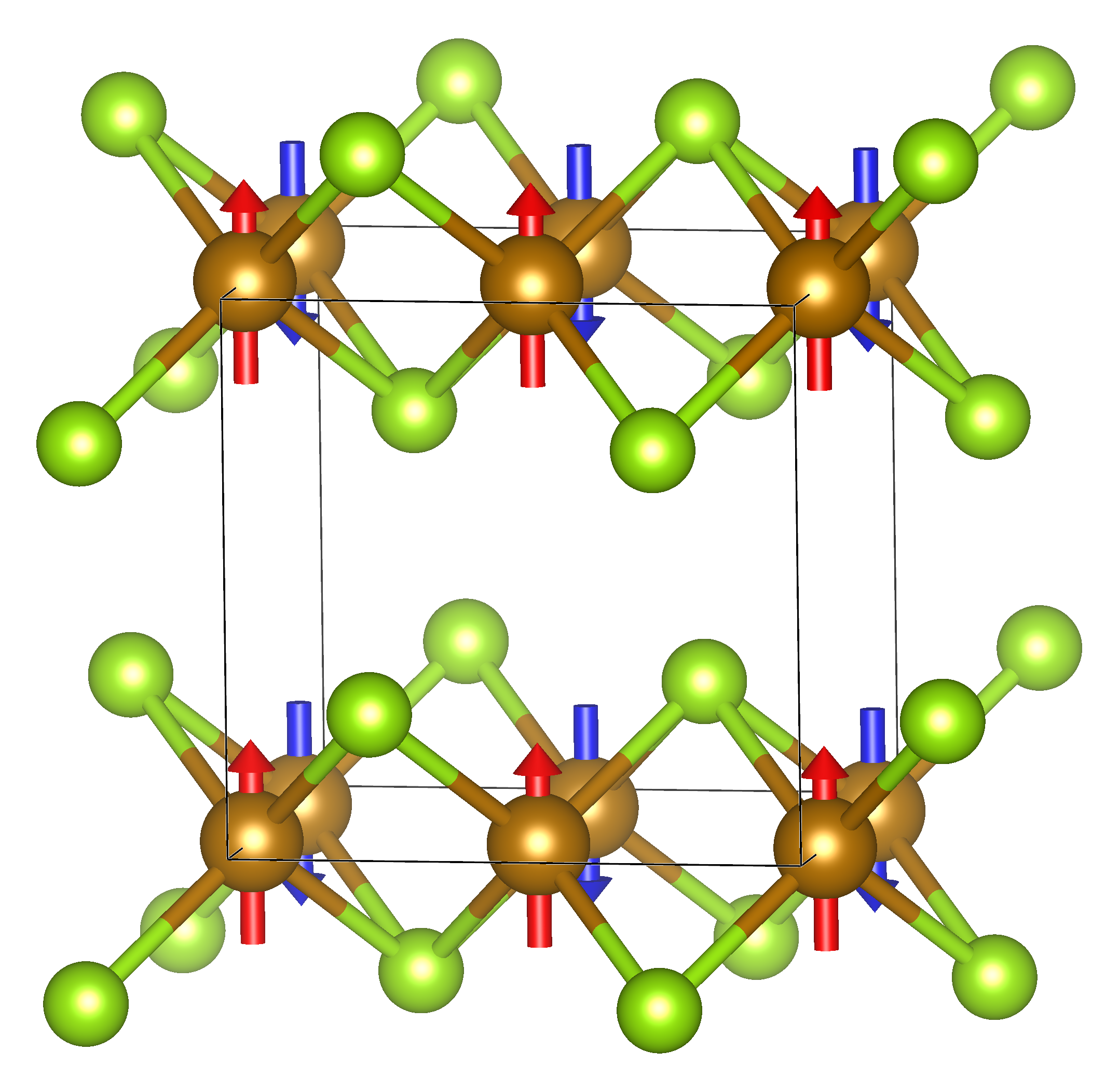}
        \caption{Stripe-2}
        \label{fig:stripe-2}
    \end{subfigure}
    \caption{FeSe magnetic unit cells for the low energy stripe-AFM configurations. (a) Stripe-0: r\textsuperscript{2}SCAN predicted ground-state configuration with inter-planar anti-parallel spin alignment. (b) Stripe-1: stripe-AFM configuration with an inter-planar alignment pattern of up-up-down-down. (c) Stripe-2: Previously reported stripe-AFM configuration with inter-planar spins aligned parallel \cite{Cao2015AntiferromagneticFeSe,Glasbrenner2015EffectChalcogenides,Liu2016NematicFeSe}. Arrows denote relative spin directions, as spin-orbit coupling was not included.}
    \label{fig:three_figures}
\end{figure*}

Figure \ref{fig:r2scan_config_energy} shows the energy difference of each configuration from the lowest energy configuration in terms of the single fixed volume calculations. The different magnetic configurations are classified into three types: AFM, spin-flipping (SF), and ferromagnetic (FM). Here, the spin-flipping configurations are defined as those where the number of spin-up Fe atoms does not equal the number of spin-down Fe atoms within its supercell, except for the FM configuration. The FM configuration has the greatest energy difference of 170.5 meV/atom. The lowest energy SF configuration has an energy difference of 18.6 meV/atom at rank no. 19. Below this rank configurations are only AFM. The eight lowest energy configurations are termed stripe-0, stripe-1, stripe-2, tetramer-0, tetramer-1, tetramer-2, tetramer-3, and criss-cross stripe. Stripe-0,1,2 are shown in Figure \ref{fig:three_figures}. The tetramers and criss-cross stripe are shown in Figures \ref{fig:tetramers} and \ref{fig:criss_cross_stripe} in the appendix. The ordering of the stripe-AFM configurations differ solely in their out-of-plane spin alignment along the \textit{c}-direction. In stripe-0, the spins are antiferromagnetically aligned. Stripe-1 exhibits an up-up-down-down spin ordering, while stripe-2 features ferromagnetic spin alignment along the \textit{c}-direction. Criss-Cross stripe also differs by out-of-plane magnetic ordering, with adjacent layers rotated by 90 degrees. Also of note are the \textit{n}-mer configurations. Here, n-mer means that a set of \textit{n} adjacent spins on a line of in-plane Fe atoms are parallelly aligned with anti-parallel alignment in the perpendicular direction. However, contrary to the proposed dimer/trimer frustrated magnetism \cite{Liu2016NematicFeSe}, the r\textsuperscript{2}SCAN + rVV10 DFA predicts the dimer and trimer configurations to be much higher in energy than the ground-state configuration. In fact, the general trend is that the energy differences decrease for higher n-mer, starting with N\'eel (1-mer) followed by Dimer (2-mer), Trimer (3-mer), Tetramer (4-mer), and stripe ($\infty$-mer) at about 109, 45, 34, 4, and 0 meV/atom, respectively.

\begin{figure}
    \centering
    \includegraphics[width=0.5\linewidth]{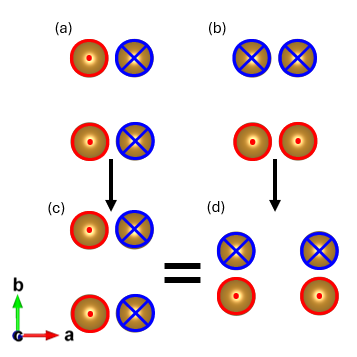}
    \caption{The example input structures, (a) and (b), relax to symmetrically equivalent output structures, (c) and (d). Note that (c) and (d) differ only by a ridged rotation. The \textit{a-b} ratios are exaggerated for demonstration. For clarity, only the spin-up Fe atoms, marked by the red $\odot$ symbols, and the spin-down Fe atoms, marked by the blue $\otimes$ symbols, in a single plane are shown. $\odot$ and $\otimes$ symbols denote relative spin directions, as spin-orbit coupling was not included.}
    \label{fig:equivalent_structures}
\end{figure}

 After the single fixed volume relaxations, it was determined that pairs of the generated symmetry-broken structures would relax to the same configuration. The process by which this happens is illustrated in Figure \ref{fig:equivalent_structures}. The two input structures denoted by (a) and (b) are not symmetrically equivalent. While (a) relaxes to an output structure, (c), that is nearly identical, (b) undergoes a stretching of the \textit{a}-axis and a contraction of the \textit{b}-axis. This lattice transformation results in structure (d), which is symmetrically equivalent to structure (c), differing only by a ridged rotation. Thus, the two symmetry-broken input structures are not separate configurations. After removing one structure from each pair from the data set, there are 278 unique configurations out of the original 383.

\begin{figure}
    \centering
    \includegraphics[width=0.5\linewidth]{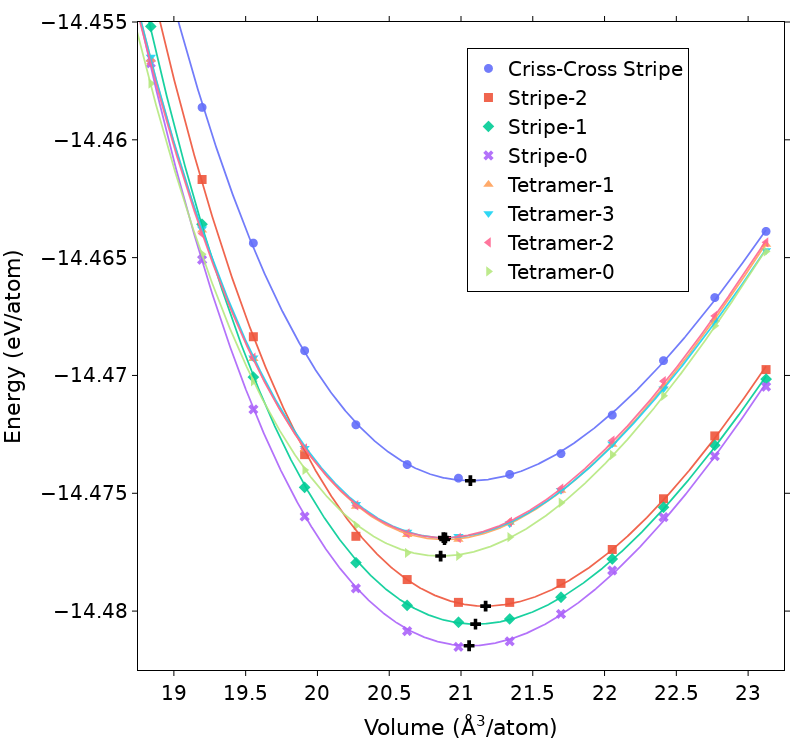}
    \caption{E-V curves predicted by r\textsuperscript{2}SCAN + rVV10 for the eight lowest unique configurations. Energy minima are marked with a cross. The Birch--Murnaghan \cite{Birch1947FiniteCrystals, Birch1978Finite300K} equation of state was used for fitting.} 
    \label{fig:r2scan_ev_curves}
\end{figure}

Figure \ref{fig:r2scan_ev_curves} and Table \ref{tab:eos_parameters} show the results of the E-V curve calculations, which more precisely determine the minimum energy for each configuration. The relative position of each configuration is maintained, and stripe-0 remains the lowest energy configuration. Energy differences for configurations with different stacking have not been reported in the literature, and to the best of our knowledge, interplanar spin coupling in FeSe has not previously been reported. However, based on the fitted EOS the energy difference from stripe-0 is $\approx$ 1.7 meV/atom for stripe-2 and $\approx$ 0.9 meV/atom for stripe-1. This is a small but significant energy difference and suggests that the paramagnetism in FeSe may be the result of magnetic frustration due to interlayer magnetic coupling. Another important feature shown is the E-V curves for the tetramer configurations, and the stripe-AFM configurations cross with the tetramers having lower energy at volumes smaller than $\SI{19}{\angstrom^3\per atom}$. This may correspond to the pressure-induced static magnetic order observed experimentally at pressures above $0.8 \si{GPa}$ \cite{Terashima2015Pressure-InducedFeSe, Bendele2012CoexistencePressure, Bendele2010PressureFeSe1-x}.

\begin{table}
    \centering
    \caption{Equilibrium properties for different configurations calculated using the Birch--Murnaghan equation of state \cite{Birch1947FiniteCrystals, Birch1978Finite300K} from the energy-volume curve first-principles calculations. $\Delta E$ (meV/atom) is the energy difference relative to the most stable configuration (Stripe-0). $V_0$ (\AA$^3$/atom) is the equilibrium volume per atom. $E_0$ (eV/atom) is the equilibrium energy per atom. $B_0$ (GPa) is the bulk modulus, representing the material's resistance to uniform compression. $B_0'$ is the first pressure derivative of the bulk modulus.}

    \label{tab:eos_parameters}
    \begin{tabular}{lccccc}
        \toprule
        Configuration & $\Delta E$ & $V_0$ & $E_0$ & $B_0$ & $B_0'$ \\
        \midrule
        Stripe-0 & 0 & 21.057 & -14.481470 & 23.85 & 8.54 \\
        Stripe-1 & 0.917 & 21.101 & -14.480553 & 23.42 & 8.57 \\
        Stripe-2 & 1.681 & 21.171 & -14.479789 & 23.54 & 7.82 \\
        Tetramer-0 & 3.806 & 20.858 & -14.477664 & 23.78 & 8.64 \\
        Tetramer-1 & 4.506 & 20.885 & -14.476964 & 23.49 & 8.69 \\
        Tetramer-2 & 4.586 & 20.877 & -14.476884 & 23.42 & 8.67 \\
        Tetramer-3 & 4.583 & 20.892 & -14.476887 & 23.17 & 8.84 \\
        Criss-Cross Stripe & 7.004 & 21.065 & -14.474467 & 23.06 & 8.70 \\
        \bottomrule
    \end{tabular}
\end{table}

Energy differences for selected configurations and a comparison to the energies reported by other first-principles calculations from the literature can be found in Table \ref{tab:common_energies}. Cao et al. \cite{Cao2015AntiferromagneticFeSe} and Glasbrenner et al. \cite{Glasbrenner2015EffectChalcogenides} both use PBE-GGA DFA, and Liu et al. \cite{Liu2016NematicFeSe} use the so-called optB86b-vdW DFA \cite{Klimes2011VanSolids}, which is also a GGA that uses an empirically optimized B86b \cite{Becke1986OnEnergy} functional for exchange, PBE for correlation, and is vdW corrected using the non-local method of \cite{Dion2004VanGeometries}. Generally, the GGA methods agree within a few meV/atom; However, the results for the r2SCAN + rVV10 are drastically different, and importantly, only the r2SCAN + rVV10 DFA yields a stripe configuration for the ground state. In contrast, the others predict the ground state to have the Dimer configuration. This result is in better agreement with inelastic neutron-scattering experiments, which show that stripe-AFM order is dominant at 4 K and low scattering energies \cite{Wang2016MagneticFeSe} although still paramagnetic, where the AFM-PM transition is currently being investigated using zentropy theory \cite{Shang2023QuantifyingFe3Pt}.

\begin{table}
    \centering
    \small
    \caption{Energy differences (meV/atom) from the ground state for selected magnetic orderings with different calculation methods. Energy differences for the stripe-AFM configurations from this work are from the more precise E-V curve calculations. The energy differences for the other listed configurations from this work are from the single fixed volume calculations.}
    \label{tab:common_energies}
    \begin{tabular}{lcccc}
        \toprule
        Magnetic Order & \makecell{r\textsuperscript{2}SCAN + rVV10 \\ (This work)} & \makecell{GGA-PBE \\ (Cao) \cite{Cao2015AntiferromagneticFeSe}} & \makecell{GGA-PBE\\ (Glasbrenner) \cite{Glasbrenner2015EffectChalcogenides}} & \makecell{optB86b-vdW \\ (Liu) \cite{Liu2016NematicFeSe}} \\
        \midrule
        Stripe-0 & 0 & - & - & - \\
        Stripe-1 & 0.9 & - & - & - \\
        Stripe-2 & 1.7 & 7.6 & 6.7 & 4.0 \\
        Dimer & 45.1 & 0 & 0 & 0 \\
        Trimer & 34.3 & - & 0.6 & 0.2 \\
        Blocked Checkerboard & 18.6 & - & 62.9 & - \\
        N\'eel & 109.0 & 21.5 & 28.9 & - \\
        Ferromagnetic & 170.5 & - & 130.2 & -\\
        \bottomrule
    \end{tabular}
\end{table}

The lattice parameters,  unit cell volumes, and magnetic moments from experiments~\cite{Kothapalli2016StrongFeSe, Horigane2009RelationshipFeSe1-xTex}, GGA-PBE~\cite{Cao2015AntiferromagneticFeSe}, GGA-PBE + rVV10L, and r\textsuperscript{2}SCAN + rVV10 are given in Table~\ref{tab:lattice_params}. The r\textsuperscript{2}SCAN + rVV10 DFA yields an Fe atomic magnetic moment of 3.08~$\mu_B$, approximately 1~$\mu_B$ larger than the GGA-PBE-based DFAs which predict Fe atomic magnetic moments around 2.0~$\mu_B$. From inelastic neutron scattering, Wang et al.~\cite{Wang2016MagneticFeSe} measured the total fluctuating magnetic moment $\langle \mu^2 \rangle = 5.19~\mu_B^2$ for FeSe at 4~K, assigning an effective spin $S_{\text{eff}} = 0.74$ using the total moment sum rule $\langle \mu^2 \rangle = (g\mu_B)^2 S(S+1)$~\cite{Lorenzana2005SumCuprates}. This result was interpreted as likely corresponding to an $S = 1$ ground state as proposed in Ref.~\cite{Wang2015NematicityFeSe}. A spin of $S = 1$ corresponds to a Fe spin magnetic moment of $2.83~\mu_B$, leading to a difference of 0.25~$\mu_B$ for r\textsuperscript{2}SCAN + rVV10 compared to 0.83~$\mu_B$ for the GGA-PBE-based DFAs. This suggests that the r\textsuperscript{2}SCAN + rVV10 DFA more accurately predicts the ground state magnetism, although more detailed consideration of excited states is required for complete accuracy at finite temperatures.

Considering the lattice parameters, the GGA-PBE + rVV10L and r\textsuperscript{2}SCAN + rVV10 calculate $c$ parameters of 5.177~\AA{} and 5.497~\AA{}, respectively, compared to the neutron powder diffraction (NPD) experimental value of 5.495~\AA{} \cite{Horigane2009RelationshipFeSe1-xTex}, resulting in differences of +318~m\AA{} and +2~m\AA{}, respectively. This suggests that the long-range vdW dispersion is accurately approximated by the r\textsuperscript{2}SCAN + rVV10 DFA, but not the GGA-PBE + rVV10L DFA. GGA-PBE + rVV10L underestimates the $a$ parameter by 0.064~\AA{} (-1.21\%) and overestimates the $b$ parameter by 0.359~\AA{} (+6.73\%) compared to high-energy X-ray diffraction (HE-XRD) measured values of $a = 5.307$~\AA{} and $b = 5.335$~\AA{} at 4~K  \cite{Kothapalli2016StrongFeSe}. The $a$ and $b$ parameters from the r\textsuperscript{2}SCAN + rVV10 DFA also misses the mark, overestimating the $a$ parameter by 0.083~\AA{} (+1.56\%) and the $b$ parameter by 0.330~\AA{} (+6.19\%). Consequently, the predicted unit cell volume from both the GGA-PBE + rVV10L and r\textsuperscript{2}SCAN + rVV10 DFAs deviate significantly from the HE-XRD experimental unit cell volume of 156~\AA{}$^3$ \cite{Kothapalli2016StrongFeSe}, with percent errors of -5.8\% and +7.7\%, respectively.

GGA-PBE without vdW correction improves the errors in the $a$ and $b$ lattice parameters, calculating $a = 5.23$~\AA{} (-1.45\% error) and $b = 5.29$~\AA{} (-0.84\% error) compared to HE-XRD \cite{Kothapalli2016StrongFeSe}. However, the $c$ parameter cannot be accurately calculated without the vdW correction due to the interlayer interactions and is not reported in this case~\cite{Cao2015AntiferromagneticFeSe}. Thus, the improvement in lattice parameters may be a coincidence, as GGA-PBE alone does not accurately account for the vdW forces in the system. It may also be that the vdW correction, while crucial for describing interlayer interactions, can degrade the accuracy of the in-plane geometry. Thus, none of the DFAs accurately determine all lattice parameters for FeSe at low temperatures. Overall, more accurate DFAs and zentropy theory \cite{Shang2023QuantifyingFe3Pt} to account for thermal effects due to low energy metastable configurations may be necessary to accurately model the lattice parameters of FeSe. Still, r\textsuperscript{2}SCAN + rVV10 yields results that better agree with experiments when considering the magnetic properties.
    
\begin{table}
    \centering
    \caption{Comparison of experimental and calculated lattice parameters, the volumes of the FeSe unit cell, and the magnetic moment of each \ch{Fe} atom $\mu_{Fe}$. \textit{c}-parameters were not reported in References \cite{Kothapalli2016StrongFeSe} and \cite{Cao2015AntiferromagneticFeSe}. HE-XRD: High-Energy X-Ray Diffraction. NPD: Neutron Powder Diffraction.}
    \label{tab:lattice_params}
    \begin{tabular}{lccccc}
        \toprule
                            & a (\AA) & b (\AA) & c (\AA) & Volume ($\text{\AA}^3$)& $\mu_{Fe}$ $(\mu_B/F)$ \\
        \midrule
        HE-XRD (4 K) \cite{Kothapalli2016StrongFeSe}  & 5.307 & 5.335 & -- & -- & -- \\
        NPD (8.5 K) \cite{Horigane2009RelationshipFeSe1-xTex}   & 5.318 & 5.343 & 5.495 & 156 & -- \\
        GGA-PBE + rVV10L (dimer)    & 5.243 & 5.694 & 5.177 & 155 & 2.00 \\
        GGA-PBE (dimer) \cite{Cao2015AntiferromagneticFeSe} & 5.23 & 5.29 & -- & -- & 2.0 \\
        r\textsuperscript{2}SCAN + rVV10 (stripe-0) & 5.390 & 5.665 & 5.497 & 168 & 3.08 \\
        
        \bottomrule
    \end{tabular}
\end{table}

Table~\ref{tab:bulk_modulus} compares the bulk modulus \(B_0\) and its first pressure derivative \(B_0'\) for two density functional approximations (DFAs) with experimental measurements. From the fitted Birch--Murnaghan equation of state (EOS), we obtain \(B_0 = 22.66\,\mathrm{GPa}\) and \(B_0' = 10.03\) for the GGA-PBE + rVV10L (dimer) configuration, and \(B_0 = 23.85\,\mathrm{GPa}\) and \(B_0' = 8.54\) for the r\textsuperscript{2}SCAN + rVV10 (stripe-0) configuration. Both values for \(B_0\) are noticeably smaller than the experimentally determined values of \(30.7(1.1)\,\mathrm{GPa}\) (XRD at 16~K) and \(33\,\mathrm{GPa}\) (NPD at 50~K)~\cite{Margadonna2009PressureK, Millican2009Pressure-inducedSuperconductor}. In addition, the computed first pressure derivatives \(B_0'\) exceed the XRD result of \(B_0' = 6.7(6)\), indicating that the DFAs predict a more rapid stiffening of the lattice under applied pressure than is observed experimentally, such a comparison however, requires zentropy for the inclusion of finite temperature effects due to the already well mixed low energy configurations in the paramagnetic phase. Further, while vdW corrections can be useful, they are not strictly part of DFT, but rather an amendment meant to account for the failures of local/semilocal DFAs to account for long range interactions. Ultimately, more accurate methods of accounting for vdW forces are required to predict the mechanical properties of FeSe.

\begin{table}
    \centering
    \caption{Comparison of experimental and calculated bulk moduli $B_0$ (GPa) and the first pressure derivative $B_0'$. XRD: X-Ray Diffraction. NPD: Neutron Powder Diffraction. The first pressure derivative was not reported for the NPD experiment.}
    \label{tab:bulk_modulus}
    \begin{tabular}{lcc}
        \toprule
            & $B_0$ & $B_0'$ \\
        \midrule
        XRD (16 K) \cite{Margadonna2009PressureK}  & 30.7(1.1) & 6.7(6) \\
        NPD (50 K) \cite{Millican2009Pressure-inducedSuperconductor} & 33 & -- \\
        GGA-PBE + rVV10L (dimer) & 22.66 & 10.03 \\
        r\textsuperscript{2}SCAN + rVV10 (stripe-0) & 23.85 & 8.54 \\
        \bottomrule
    \end{tabular}
\end{table}

\section{Summary}
The present findings suggest that the DFT-based calculations in terms of r\textsuperscript{2}SCAN + rVV10 provide a more accurate description of the magnetic ground state of FeSe, predicting a stripe-AFM configuration that aligns with experimental observations of dominant stripe fluctuations at low temperatures \cite{Wang2016MagneticFeSe}. This represents an improvement over the GGA-based density functional approximations, which predict a dimer ground state configuration inconsistent with experiments. However, the present work also highlights ongoing challenges in accurately predicting lattice parameters, the bulk modulus, and magnetic moments for FeSe, suggesting that while the r\textsuperscript{2}SCAN meta-GGA enhances predictive accuracy, further refinement or alternative approaches may be needed to achieve agreement with experimental data. Moreover, future work should employ zentropy theory \cite{Shang2023QuantifyingFe3Pt}---an approach that integrates quantum-mechanical and statistical-mechanical frameworks---to capture finite-temperature effects and the complex interplay of low-energy magnetic states. Finally, these results underscore the importance of using accurate exchange-correlation functionals in studying complex magnetic materials and contribute to the broader understanding of the magnetic properties of iron-based superconductors.

\section{Acknowledgments}
The present work is funded by the Department of Energy (DOE) through Grant No. DE-SC0023185 and used the Roar Collab cluster at the Pennsylvania State University and the Extreme Science and Engineering Discovery Environment (XSEDE), which is supported by National Science Foundation grant number ACI-1548562. Specifically, it used the Bridges-2 system, which is supported by NSF award number ACI-1928147, at the Pittsburgh Supercomputing Center (PSC).

\printbibliography

\newpage
\appendix
\renewcommand\thefigure{\thesection.\arabic{figure}}
\setcounter{figure}{0}
\renewcommand{\thetable}{A\arabic{table}}
\setcounter{table}{0}
\section{Appendix}

\begin{table}[h!]
    \caption{VASP input settings for the single fixed volume and E-V curve calculations. The values in parentheses indicate settings for final energy calculation in the three-step relaxation process for each point when different. ENCUT: energy cutoff for the plane-wave basis set. EDIFF: The global break condition for the electronic self-consistency loop. KPPA: k-points per reciprocal atom. IBRION: Structure optimization procedure. ISMEAR: partial occupancy smearing. Input Fe MAGMOM: initial atomic magnetic moment for iron atoms.}
    \label{tab:vasp_settings}
    \centering
    \begin{tabular}{ccc}
        \toprule
        Setting & Single Fixed Volume & EV-Curve \\
        \midrule
        ENCUT & 520 eV & 520 eV \\
        EDIFF & $2 \times 10^{-4} \ \si{eV}$ & $4 \times 10^{-7} \ \si{eV}$ \\
        KPPA & 6000 & 8000 \\
        IBRION & 2 & 2 (-1) \\
        ISMEAR & 0 & 0 (-5) \\
        Input Fe MAGMOM & $\pm 5 \ \si{\mu_B}$ & $\pm 5 \ \si{\mu_B}$ \\
        Volume  & 21.5 $(\text{\AA}^3/\text{atom})$  & 18.125 - 23.125 $(\text{\AA}^3/\text{atom})$ \\
        \bottomrule
    \end{tabular}
\end{table}

\begin{figure}[h!]
    \centering
    \includegraphics[width=0.5\linewidth]{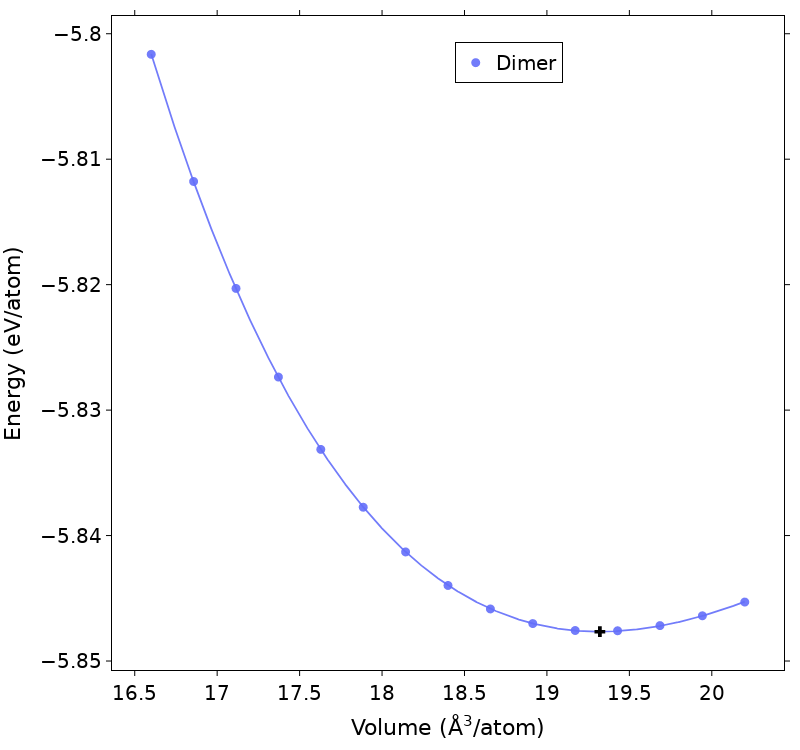}
    \caption{PBE + rVV10L energy-volume curve for the Dimer configuration. The Birch--Murnaghan equation of state was used for fitting \cite{Birch1947FiniteCrystals, Birch1978Finite300K}.}
    \label{fig:pbe_rvv10l_dimer_ev}
\end{figure}

\begin{figure}
    \centering
    \includegraphics[width=0.5\linewidth]{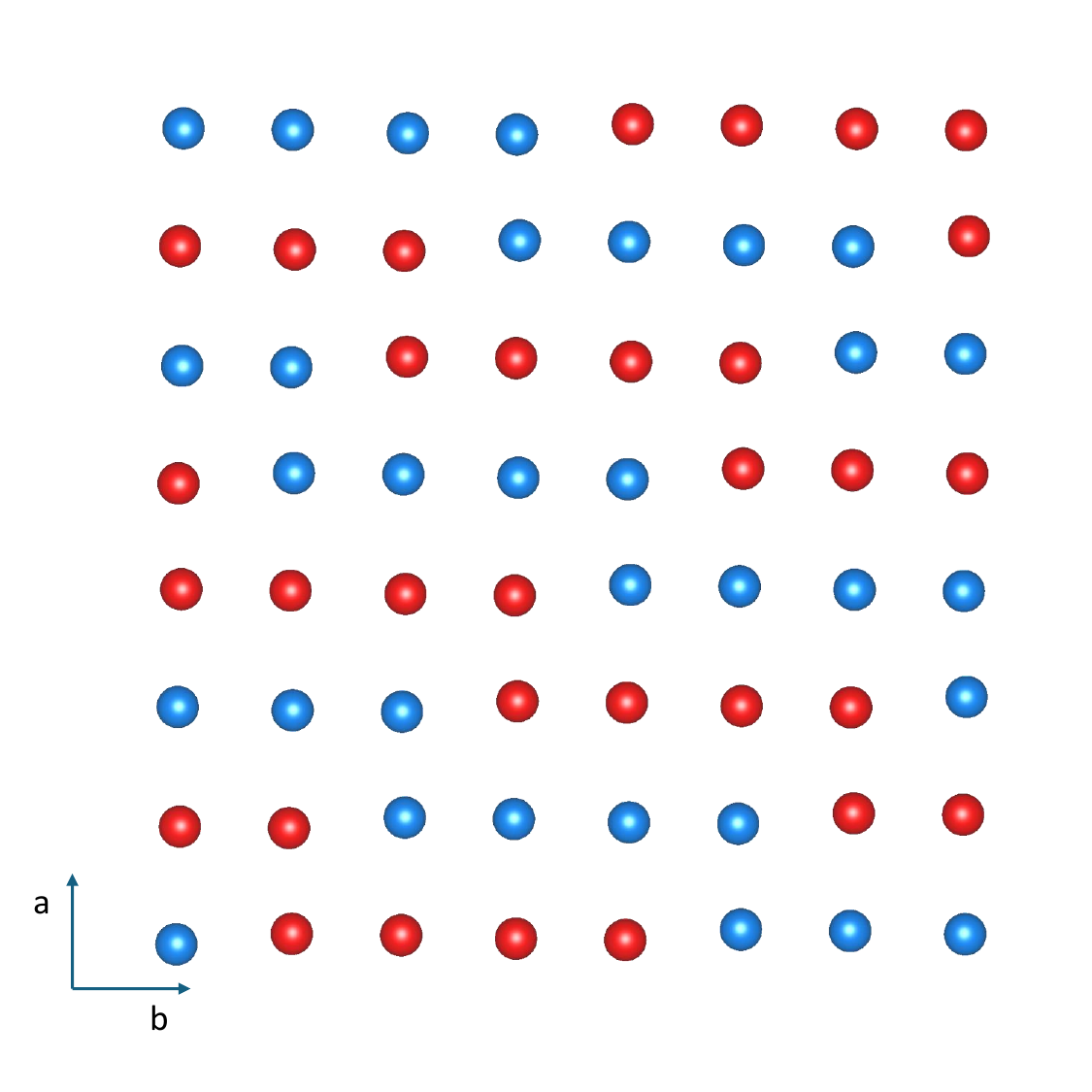}
    \caption{a-b layer of the tetramer configurations. Spin up (down) atoms are shown in red (blue). Tetramer-0 has up-up-down-down interlayer stacking. Tetramer-1 has up-up-down interlayer stacking. Tetramer-2 has antiferromagnetic interlayer stacking. Tetramer-3 has ferromagnetic interlayer stacking. Note that the primitive cell for each can be made with 16 atoms or less, as the primitive lattice vectors need not be orthogonal. See poscars and incars in the Ancillary files for the primitive cells.}
    \label{fig:tetramers}
\end{figure}

\begin{figure*}
    \centering
    \begin{subfigure}[t]{0.14\textwidth}
        \centering
        \includegraphics[width=1\linewidth]{compass.png}
    \end{subfigure}%
    ~
    \begin{subfigure}[t]{0.75\textwidth}
        \centering
        \includegraphics[width=1\linewidth]{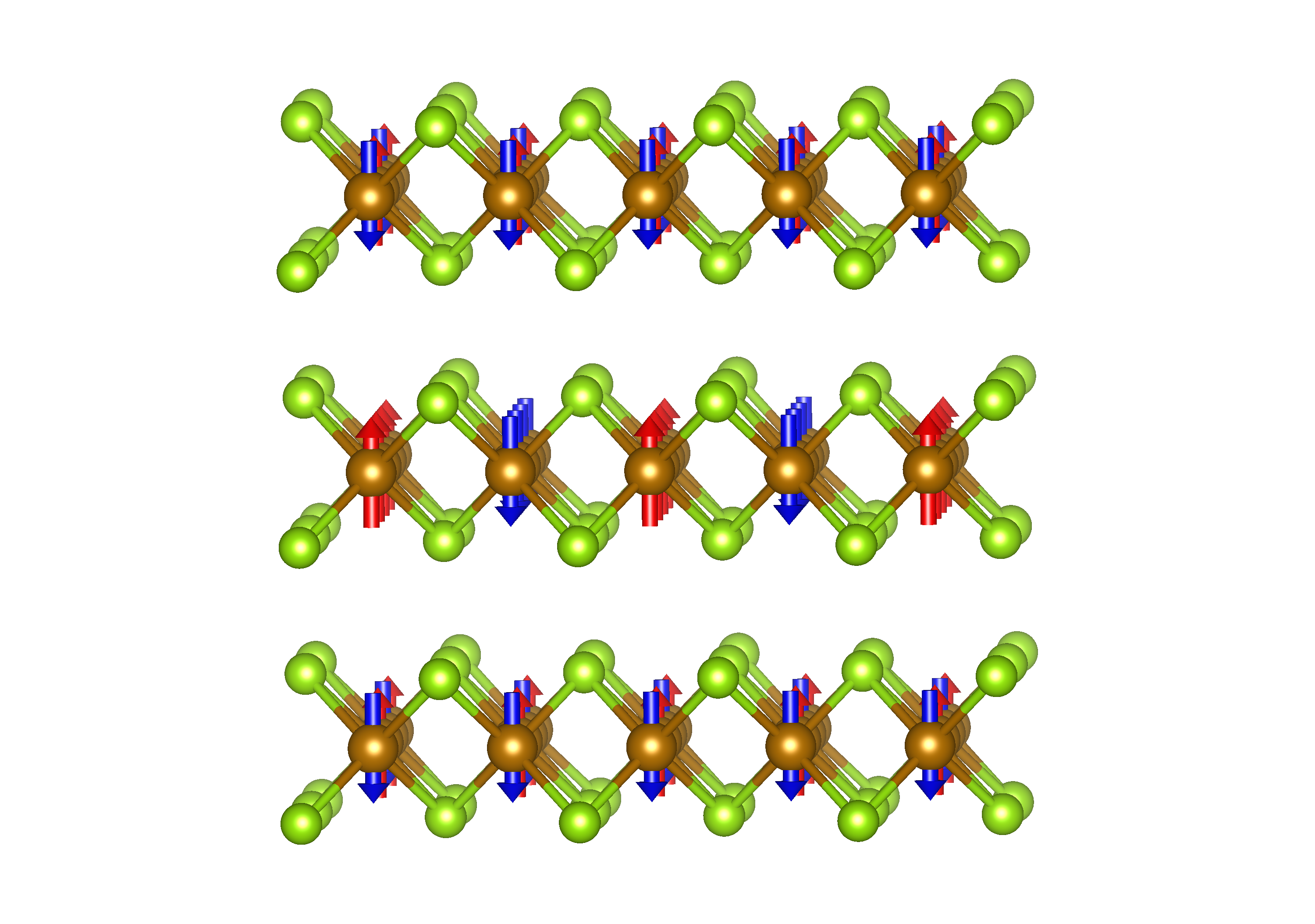}
    \end{subfigure}%
    \caption{Criss-cross stripe configuration. Adjacent layers differ by a 90 degree rotation.}
    \label{fig:criss_cross_stripe}
\end{figure*}

\end{document}